\documentclass[amsmath,amssymb]{revtex4}
\usepackage{graphicx}% Include figure files
\usepackage{dcolumn}% Align table columns on decimal point
\usepackage{bm}% bold math

\newcommand{\gsim}{\raisebox{0.2ex}{$\ > \kern -1.05em%
        \raisebox{-1.1ex}{$\sim$}\ $}}  % >=
\begin{document}

\title{A Formulation of the Ring Polymer Molecular Dynamics
}

\author{Atsushi Horikoshi} 
\email{horikosi@tcu.ac.jp}
\affiliation{
Department of Natural Sciences, Faculty of Knowledge Engineering, Tokyo City University, 
Tamazutsumi, Setagaya-ku, Tokyo 158-8557, Japan
}

\begin{abstract}
The exact formulation of the path integral centroid dynamics 
is extended to include composites of 
the position and momentum operators.
We present the generalized centroid dynamics (GCD),
which provides a basis to calculate 
Kubo-transformed correlation functions 
by means of classical averages.
We define various types of approximate GCD,
one of which is equivalent to 
the ring polymer molecular dynamics (RPMD).
The RPMD and another approximate GCD are 
tested in one-dimensional harmonic system, 
and it is shown that the RPMD works better   
in the short time region.

\end{abstract}

\maketitle

\section{Introduction}
\hspace*{\parindent}
There has been great interest in revealing 
quantum dynamical aspects of condensed-phase molecular systems
such as the liquid hydrogen and liquid helium.
The path integral formulation of quantum mechanics \cite{FeynmanHibbs}
is most suited for numerical analysis of complex molecular systems
to provide the basis of 
the path integral Monte Carlo (PIMC)
and path integral molecular dynamics (PIMD) 
\cite{Ceperley,BerneThirumalai,BerneCiccottiCoker}.
Most of the static equilibrium properties of finite temperature quantum systems
can be computed by means of the PIMC/PIMD techniques.
However, it is difficult to apply the PIMC/PIMD directly 
to compute dynamical properties such as real time quantum correlation functions.
This is because the imaginary time path integral formalism
is used in the PIMC/PIMD, in which 
we exploit the isomorphism between the imaginary time path integral
representation of the quantum partition function and 
the classical partition function of a fictitious ring polymer 
\cite{Ceperley,BerneThirumalai,BerneCiccottiCoker}.
In the PIMD simulations, physical quantities are evaluated by time averages of the
dynamical variables as in ordinary classical molecular dynamics simulations.
However, the time evolution of each bead $q_{j}$
in the ring polymer is just fictitious,
and, therefore, we cannot directly evaluate real time-dependent properties by means
of the PIMD method.
\\
\hspace*{\parindent}
A number of quantum dynamics methods to calculate 
real time quantum correlation functions
have been proposed so far
\cite{Miller,RabaniReichmanKrilovBerne,RabaniReichman}.
In this article, we focus on quantum dynamics methods to 
calculate Kubo-transformed
correlation functions \cite{KuboTodaHashitsume}
$\langle \hat{B}(0) \hat{A}(t)\rangle^{\rm K}_{\beta}$ 
by means of the PIMD technique. 
Kubo-transformed correlation functions are important quantities that characterize 
dynamical effects in quantum mechanical systems, and  
play a central role in the linear response theory \cite{KuboTodaHashitsume}.
Full quantum correlation functions $\langle \hat{B}(0) \hat{A}(t)\rangle_{\beta}$
can be reproduced from the Kubo-transformed correlation functions
using the relation in the frequency space: 
$\langle \hat{B}\hat{A}\rangle_{\beta}(\omega)=E(\omega)
\langle \hat{B}\hat{A}\rangle^{\rm K}_{\beta}(\omega)$, where
$E(\omega)$ is a known factor \cite{KuboTodaHashitsume}.
\\
\hspace*{\parindent}
Two methods have been proposed so far.
One is the centroid molecular dynamics (CMD) \cite{CaoVoth}
and the other is the ring polymer molecular dynamics (RPMD) \cite{CraigManolopoulos}.
The CMD is a classical dynamics of the centroid variables 
$(q_{c},p_{c})$ on the effective classical 
potential surface \cite{FeynmanKleinert,GiachettiTognetti}.
The name ``CMD'' comes from the fact that the variable $q_{c}$ corresponds 
to the ``centroid'' of the ring polymer beads: 
$q_{0}=(1/P)\sum_{j=1}^{P}q_{j}$.
The effective classical potential is a quantum mechanically corrected potential,
and can be evaluated by PIMC, PIMD, or other methods.
In the most of CMD simulations, however,
the ``on the fly'' integration scheme \cite{CaoVoth4} is employed, 
and, therefore, the CMD is usually implemented by means of the PIMD technique.   
The CMD can be derived from a more fundamental dynamics, the centroid dynamics (CD),
which is defined by the quasi-density operator (QDO) formalism 
for operators ($\hat{q},\hat{p}$) \cite{JangVoth,JangVoth2}.
It has been shown that a correlation function given by the CD is equivalent 
to the corresponding Kubo-transformed correlation function 
if the operator $\hat{B}$ is linear in $\hat{q}$ and $\hat{p}$ \cite{JangVoth}.     
Therefore, the CD and CMD are not applicable to the case that
$\hat{B}$ is nonlinear in them.
This is the nonlinear operator problem in the CD.
This difficulty can be avoided by considering an improved CD, correlation functions
given by which correspond to the higher order Kubo-transformed
correlation functions \cite{ReichmanRoyJangVoth}. 
However, it is in general difficult to convert those correlation functions to 
$\langle \hat{B}(0) \hat{A}(t)\rangle_{\beta}$ or  
$\langle \hat{B}(0) \hat{A}(t)\rangle_{\beta}^{\rm K}$.
\\
\hspace*{\parindent}
On the other hand, the RPMD is a quite simple method.
We just identify the fictitious time evolution 
of the ring polymer beads in the PIMD 
as the real time evolution,
and then calculate correlation functions 
$\langle B_{0}(0) A_{0}(t)\rangle_{\beta}^{\rm RPMD}$,
where $A_{0}$ and $B_{0}$ are  
the centroids of position-dependent operators 
$A(\hat{q})$ and $B(\hat{q})$, respectively \cite{CraigManolopoulos}.
The RPMD has some advantages: 
we can treat operators nonlinear in $\hat{q}$, 
and it works well in the short time region to be exact 
at $t=0$ \cite{BraamsManolopoulos}.
These good properties are ensured by the fact that 
an identity between a static Kubo-transformed correlation function 
and the corresponding quantity defined by the path integral,
$\langle \hat{B}(0) \hat{A}(0)\rangle_{\beta}^{\rm K}$=
$\langle B_{0}(0) A_{0}(0)\rangle_{\beta}^{\rm PI}$, 
holds for any position-dependent operators. 
The RPMD looks promising, however, this is just a model 
and has not been derived from more fundamental theories
so far. 
\\
\hspace*{\parindent}
Recently, Hone {\it et al.} have pointed out that 
for the case of $\hat{A}=\hat{B}=\hat{q}$,
the RPMD is equivalent to the CMD approximated by 
the instantaneous force approximation \cite{HoneRosskyVoth}.
Their observation is quite suggestive.
This correspondence implies that 
the RPMD can be formulated as an approximate dynamics 
of a more fundamental dynamics, 
just like the CMD is formulated as an approximate CD. 
However, there is a problem here: 
The correspondence found by them is limited for the case of 
$\hat{A}=\hat{B}=\hat{q}$. 
This is because the CD and CMD have the nonlinear operator problem.
Therefore, the QDO formalism should be extended
to include operators nonlinear in $\hat{q}$.
\\
\hspace*{\parindent}
In this article, we develop a new QDO formalism which includes 
composite operators of $\hat{q}$ and $\hat{p}$.
A new CD, the generalized centroid dynamics (GCD), is
defined by means of this formalism.
We then show that one of the approximate GCD is equivalent to 
the RPMD.
\\
\hspace*{\parindent}
This article is organized as follows.
In Sec. 2, we introduce the effective classical potential and density
for composite operators. 
In Sec. 3, the QDO formalism is extended to include composite operators,
and the GCD is then defined.
In Sec. 4, several approximate dynamics are presented and 
tested in a simple system.     
Conclusions are given in Sec. 5.

%%%%%%%%%%%%%%%%%%%%%%%%%%%%%%%%%%%%%%%%%%%%%%%%%%%%%%%%%%%%%%%%%%%%%%%
\section{Effective classical potentials and effective classical densities 
for composite operators}
%%%%%%%%%%%%%%%%%%%%%%%%%%%%%%%%%%%%%%%%%%%%%%%
\subsection{Effective classical potentials for composite operators}
%%%%%%%%%%%%%%%%%%%%%%%%%%%%%%%%%%%%%%%%%%%%%%%
\hspace*{\parindent}
Consider a quantum system, the Hamiltonian of which is given by
\begin{eqnarray}
\hat{H}=\frac{1}{2m}\hat{p}^{2}+V(\hat{q}).
\label{1}
\end{eqnarray}
For simplicity, we use one-dimensional notation in this article,
but the multidimensional generalization is straightforward. 
The density operator of this system is $\hat{\rho}_{\beta}=e^{-\beta\hat{H}}$
and the quantum mechanical partition function is given by the trace of it, 
$Z_{\beta}={\rm Tr}~\hat{\rho}_{\beta}$; 
here, $\beta=1/(k_{B}T)$ is the inverse temperature.
The phase space path integral representation of $Z_{\beta}$ is written as
\cite{Kleinert}
\begin{eqnarray}
Z_{\beta}&=&\int^{\infty}_{-\infty}\!\!\!dq\int\!\!\!\!\int^{q(\beta\hbar)=q}_{q(0)=q}
\!\!\!{\cal D}q{\cal D}p~e^{-S[q,p]/\hbar}
\label{2a}\\
&=&\lim_{P\to\infty}
\prod_{j=1}^{P}\int\!\!\!\!\int\frac{dq_{j}dp_{j}}{2\pi\hbar}~
e^{-\beta_{P}H_{P}^{\rm ps}},
\label{2b}
\end{eqnarray}
where the imaginary time interval $[0,\beta\hbar]$ is discretized into
$P$ slices, $q_{j}$ and $p_{j}$ are  
the position and momentum at the $j$-th time slice, respectively,
and $\beta_{P}=\beta/P$. 
The action $S[q,p]$ is defined by 
\begin{eqnarray}
S[q,p]=\int^{\beta\hbar}_{0}\!\!\!d\tau\left(H[q,p]-ip\dot{q}\right),
\label{3}
\end{eqnarray}
and the Hamiltonian of $2P$ variables $\{q_{j},p_{j}\}$ is defined by
\begin{eqnarray}
H_{P}^{\rm ps}(\{q_{j},p_{j}\})=\sum_{j=1}^{P}\left.\left[
\frac{p^{2}_{j}}{2m}+V(q_{j})-i\frac{p_{j}(q_{j}-q_{j-1})}{\beta_{P}\hbar}
\right]\right|_{q_{0}=q_{P}}.
\label{4}
\end{eqnarray}
\\
\hspace*{\parindent}
Consider position-dependent Hermitian operators
$\{\hat{O}^{(i)}=O^{(i)}(\hat{q})\}~(i=1\sim N)$
which are given by operator products of $\hat{q}$ or sums of them.
We refer to these operators as composite operators.
For each operator $\hat{O}^{(i)}$,
there exists a corresponding ``centroid'' $O_{0}^{(i)}$
in the path integral representation,
\begin{eqnarray}
O_{0}^{(i)}=
\frac{1}{\beta\hbar}\int^{\beta\hbar}_{0}\!\!\!d\tau~O^{(i)}(q(\tau))
=\lim_{P\to\infty}\frac{1}{P}
\sum_{j=1}^{P}O^{(i)}(q_{j}).
\label{5}
\end{eqnarray}
Inserting $N$ identities $1=\int^{\infty}_{-\infty}dO_{c}^{(i)}
\delta(O^{(i)}_{0}-O^{(i)}_{c})$ into 
the partition function $Z_{\beta}$ (Eq. (\ref{2a})), 
we rewrite the partition function as
\begin{eqnarray}
Z_{\beta}=C\prod_{i=1}^{N}\int^{\infty}_{-\infty}\!\!\!dO_{c}^{(i)}~
e^{-\beta U_{\beta}^{c}(\{O_{c}^{(i)}\})},
\label{6}
\end{eqnarray}
where $O_{c}^{(i)}$ is the static centroid variable that corresponds to
the centroid $O_{0}^{(i)}$,
and 
$U_{\beta}^{c}(\{O_{c}^{(i)}\})$ is the effective classical potential
for composite operators
\begin{eqnarray}
U_{\beta}^{c}(\{O_{c}^{(i)}\})=-\frac{1}{\beta}\log\left[\int^{\infty}_{-\infty}
\!\!\!dq\int\!\!\!\!\int^{q(\beta\hbar)=q}_{q(0)=q}\!\!\!{\cal D}q{\cal D}p~
\prod_{i=1}^{N}\delta(O^{(i)}_{0}-O^{(i)}_{c})e^{-S[q,p]/\hbar}\right]+C.
\label{7}
\end{eqnarray}
Here and hereafter, we represent irrelevant constant factors by a symbol $C$. 
This is the constraint effective potential with many constraints on
composite operators, which has been introduced by Fukuda and Kyriakopoulos
in the context of quantum field theories \cite{FukudaKyriakopoulos}.
%%%%%%%%%%%%%%%%%%%%%%%%%%%%%%%%%%%%%%%%%%%%%%%
\subsection{Effective classical densities for composite operators}
%%%%%%%%%%%%%%%%%%%%%%%%%%%%%%%%%%%%%%%%%%%%%%%
\hspace*{\parindent}
Next, consider Hermitian composite operators 
$\{\hat{O}^{(i)}={O}^{(i)}(\hat{q},\hat{p})\}~(i=1\sim N)$
which are given by operator products of $\hat{q}$,
operator products of $\hat{p}$, or sums of them. 
It should be noted that operators which 
are products of $\hat{q}$ and $\hat{p}$, 
such as $(\hat{q}\hat{p}+\hat{p}\hat{q})/2$,
are excluded here. 
The corresponding centroids are given as
\begin{eqnarray}
O_{0}^{(i)}=\frac{1}{\beta\hbar}\int^{\beta\hbar}_{0}
\!\!\!d\tau~O^{(i)}(q(\tau),p(\tau))
=\lim_{P\to\infty}\frac{1}{P}\sum_{j=1}^{P}O^{(i)}(q_{j},p_{j}).
\label{8}
\end{eqnarray}
Inserting $N$ identities $1=\int^{\infty}_{-\infty}dO_{c}^{(i)}
\delta(O^{(i)}_{0}-O^{(i)}_{c})$ into Eq. (\ref{2a}), we obtain another expression 
of the quantum partition function 
\begin{eqnarray}
Z_{\beta}=C\prod_{i=1}^{N}\int^{\infty}_{-\infty}\!\!\!dO_{c}^{(i)}~
\rho_{\beta}^{c}(\{O_{c}^{(i)}\}),
\label{9}
\end{eqnarray}
where 
\begin{eqnarray}
\rho_{\beta}^{c}(\{O_{c}^{(i)}\})&=&\frac{1}{C}\int^{\infty}_{-\infty}
\!\!\!dq\int\!\!\!\!\int^{q(\beta\hbar)=q}_{q(0)=q}\!\!\!{\cal D}q{\cal D}p~
\prod_{i=1}^{N}\delta(O^{(i)}_{0}-O^{(i)}_{c})e^{-S[q,p]/\hbar}
\label{10a}\\
&=&e^{-\beta H_{\beta}^{c}(\{O_{c}^{(i)}\})}
\label{10b}
\end{eqnarray}
is the effective classical density for composite operators, 
and $H_{\beta}^{c}(\{O_{c}^{(i)}\})$ is the effective classical Hamiltonian.
These effective classical quantities enable a classical description 
of static quantum properties.
If we choose $N=2$ and $(\hat{O}^{(1)},\hat{O}^{(2)})=(\hat{q},\hat{p})$,
the effective classical Hamiltonian becomes the ordinary one \cite{JangVoth}
\begin{eqnarray}
H_{\beta}^{c}(q_{c},p_{c})=
\frac{p_{c}^{2}}{2m}+V_{\beta}^{c}(q_{c}),
\label{11}
\end{eqnarray}
where $V_{\beta}^{c}(q_{c})$ is the ordinary effective classical potential
\cite{FeynmanKleinert,GiachettiTognetti}.

%%%%%%%%%%%%%%%%%%%%%%%%%%%%%%%%%%%%%%%%%%%%%%%%%%%%%%%%%%%%%%%%%%%%%%%
\section{Exact centroid dynamics for composite operators}\label{GCD}
\hspace*{\parindent}
In this section, we extend the QDO formalism \cite{JangVoth}
to include Hermitian composite operators 
$\{\hat{O}^{(i)}=O^{(i)}(\hat{q},\hat{p})\}$ 
introduced in the preceding section.
Then, we define an exact dynamics of centroid variables 
and show the exact correspondence between 
centroid correlation functions and  
Kubo-transformed correlation functions. 

%%%%%%%%%%%%%%%%%%%%%%%%%%%%%%%%%%%%%%%%%%%%%%%
\subsection{QDO for composite operators}
%%%%%%%%%%%%%%%%%%%%%%%%%%%%%%%%%%%%%%%%%%%%%%%
\hspace*{\parindent}
The canonical density operator for the Hamiltonian (Eq. (\ref{1})) 
can be decomposed as
\begin{eqnarray}
\hat{\rho}_{\beta}=C\prod_{i=1}^{N}\int^{\infty}_{-\infty}\!\!\!dO_{c}^{(i)}~
\hat{\varphi}^{c}_{\beta}(\{O_{c}^{(i)}\}),
\label{12}
\end{eqnarray}
where we introduced an operator
\begin{eqnarray}
\hat{\varphi}_{\beta}^{c}(\{O_{c}^{(i)}\})=
\frac{1}{C}\prod_{i=1}^{N}\int^{\infty}_{-\infty}\!\frac{d\eta_{i}}{2\pi}~
e^{-\beta \hat{H}+i\sum_{i=1}^{N}\eta_{i}(\hat{O}^{(i)}-O_{c}^{(i)})}.
\label{13}
\end{eqnarray}
One can see the validity of this decomposition as follows.
A position space matrix element of this operator has
its phase space path integral representation,
\begin{eqnarray}
\langle q_{b}|\hat{\varphi}_{\beta}^{c}(\{O_{c}^{(i)}\})|q_{a}\rangle
=\frac{1}{C}\int\!\!\!\!\int^{q(\beta\hbar)=q_{b}}_{q(0)=q_{a}}\!{\cal D}q{\cal D}p~
\prod_{i=1}^{N}\int^{\infty}_{-\infty}\!\frac{d\eta_{i}}{2\pi}~
e^{i\eta_{i}(O^{(i)}_{0}-O^{(i)}_{c})}
e^{-S[q,p]/\hbar}.
\label{14}
\end{eqnarray}
Taking the trace of this matrix element and using the integral expression 
of the $\delta$ function, $(2\pi)\delta(O_{0}-O_{c})=
\int^{\infty}_{-\infty}\!d\eta ~e^{i\eta(O_{0}-O{c})}$, we reproduce 
the effective classical density for composite operators (Eq. (\ref{10a}))
\begin{eqnarray}
\rho^{c}_{\beta}(\{O_{c}^{(i)}\})=
{\rm Tr}\left[\hat{\varphi}_{\beta}^{c}(\{O_{c}^{(i)}\})\right].
\label{15}
\end{eqnarray}
The quantum partition function $Z_{\beta}$ is then reproduced 
using the expression (Eq. (\ref{9})).
We define the QDO
for composite operators by normalizing the operator $\hat{\varphi}_{\beta}^{c}$,
\begin{eqnarray}
\hat{\delta}_{\beta}^{c}(\{O_{c}^{(i)}\})=
\frac{\hat{\varphi}_{\beta}^{c}(\{O_{c}^{(i)}\})}{\rho_{\beta}^{c}(\{O_{c}^{(i)}\})}.
\label{16}
\end{eqnarray}
%%%%%%%%%%%%%%%%%%%%%%%%%%%%%%%%%%%%%%%%%%%%%%%
\subsection{Generalized centroid dynamics}
%%%%%%%%%%%%%%%%%%%%%%%%%%%%%%%%%%%%%%%%%%%%%%%
\hspace*{\parindent}
We define an exact time evolution of the QDO as
\begin{eqnarray}
\hat{\delta}_{\beta}^{c}(t;\{O_{c}^{(i)}\})=
e^{-i\hat{H}t/\hbar}~\hat{\delta}_{\beta}^{c}(\{O_{c}^{(i)}\})~e^{i\hat{H}t/\hbar}.
\label{18b}
\end{eqnarray}
A dynamical centroid variable $O_{c}^{(i)}(t)$ is then defined by 
\begin{eqnarray}
O_{c}^{(i)}(t)=
{\rm Tr}\left[\hat{\delta}_{\beta}^{c}(t;\{O_{c}^{(i)}\})\hat{O}^{(i)}\right]=
{\rm Tr}\left[\hat{\delta}_{\beta}^{c}(\{O_{c}^{(i)}\})\hat{O}^{(i)}(t)\right].
\label{18c}
\end{eqnarray}
This is the GCD, an exact CD 
for composite operators.
In the case of $N=2$ and $(\hat{O}^{(1)},\hat{O}^{(2)})=(\hat{q},\hat{p})$,
the GCD is reduced to the original CD
proposed by Jang and Voth \cite{JangVoth}.
\\
\hspace*{\parindent}
Next, consider a Hermitian operator 
$\hat{A}=A(\hat{q},\hat{p})$
which is given by operator products of $\hat{q}$,
operator products of $\hat{p}$, or sums of them. 
In terms of the QDO, a classical counterpart to the operator $\hat{A}$ 
is defined as
\begin{eqnarray}
A_{\beta}^{c}(\{O_{c}^{(i)}\})=
{\rm Tr}\left[\hat{\delta}_{\beta}^{c}(\{O_{c}^{(i)}\})\hat{A}\right].
\label{17}
\end{eqnarray}
This is the effective classical operator for $\hat{A}$, 
which is a function 
of static centroid variables $\{O_{c}^{(i)}\}$.
The time dependent effective classical operator
is given by using time dependent QDO (Eq. (\ref{18b})),
\begin{eqnarray}
A_{\beta}^{c}(t)=A_{\beta}^{c}(t;\{O_{c}^{(i)}\})=
{\rm Tr}\left[\hat{\delta}_{\beta}^{c}(t;\{O_{c}^{(i)}\})\hat{A}\right]=
{\rm Tr}\left[\hat{\delta}_{\beta}^{c}(\{O_{c}^{(i)}\})\hat{A}(t)\right].
\label{18a}
\end{eqnarray}
If the operator $\hat{A}$ belongs to the operator set $\{\hat{O}^{(i)}\}$
which is chosen to define the QDO, the effective classical operator 
is equal to the static centroid variable, 
$A_{\beta}^{c}(\{O_{c}^{(i)}\})=A_{c}$,
and the time-dependent effective classical operator equals
the dynamical centroid variable
$A_{\beta}^{c}(t,\{O_{c}^{(i)}\})=A_{c}(t)$.
%%%%%%%%%%%%%%%%%%%%%%%%%%%%%%%%%%%%%%%%%%%%%%%
\subsection{Correlation functions}
%%%%%%%%%%%%%%%%%%%%%%%%%%%%%%%%%%%%%%%%%%%%%%%
\hspace*{\parindent}
Consider a couple of Hermitian composite operators, 
$\hat{A}=A(\hat{q},\hat{p})$ and $\hat{B}=B(\hat{q},\hat{p})$.
In terms of the corresponding effective classical operators, we define
the centroid correlation function by means of a ``classical'' ensemble average,
\begin{eqnarray}
\langle B^{c}_{\beta}(0)A^{c}_{\beta}(t)\rangle^{\rm CD}_{\beta}&=&
\frac{1}{Z_{\beta}}
C\prod_{i=1}^{N}\int^{\infty}_{-\infty}\!\!\!dO_{c}^{(i)}~
\rho_{\beta}^{c}(\{O_{c}^{(i)}\})
~B_{\beta}^{c}A_{\beta}^{c}(t).
\label{20}
\end{eqnarray}
If the operator $\hat{B}$ belongs to the operator set $\{\hat{O}^{(i)}\}$, 
the centroid correlation function (Eq. (\ref{20})) is identical 
to the Kubo-transformed correlation function 
\cite{KuboTodaHashitsume,JangVoth}:
\begin{eqnarray}
\langle B^{c}_{\beta}(0)A^{c}_{\beta}(t)\rangle^{\rm CD}_{\beta}
&=&
\langle B_{c}(0)A^{c}_{\beta}(t)\rangle^{\rm CD}_{\beta}\nonumber\\
&=&\frac{1}{Z_{\beta}}\int^{\beta}_{0}\!\!\frac{d\lambda}{\beta}~
{\rm Tr}\left[e^{-(\beta-\lambda)\hat{H}}\hat{B}~
e^{-\lambda\hat{H}}\hat{A}(t)\right]\nonumber\\
&=&\langle \hat{B}(0)\hat{A}(t)\rangle^{\rm K}_{\beta},
\label{21}
\end{eqnarray} 
This identity gives us an effective classical way to calculate 
Kubo-transformed correlation functions.
The GCD procedure to calculate $\langle \hat{B}(0)\hat{A}(t)\rangle^{\rm K}_{\beta}$
is summarized as follows:
First, we choose an operator set $\{\hat{O}^{(i)}\}$ 
which includes the operator $\hat{B}$,
and calculate the effective classical density 
$\rho_{\beta}^{c}(\{O_{c}^{(i)}\})$.
Second, we compute the GCD trajectories $A_{\beta}^{c}(t)$
by evolving $A_{\beta}^{c}(\{O_{c}^{(i)}\})$,
where the initial values of the centroid variables $\{O_{c}^{(i)}\}$
are given by the distribution $\rho_{\beta}^{c}(\{O_{c}^{(i)}\})$.
Averaging over the GCD trajectories,
we obtain the centroid correlation function (Eq. (\ref{20})),
which is identical to the Kubo-transformed correlation function.
%%%%%%%%%%%%%%%%%%%%%%%%%%%%%%%%%%%%%%%%%%%%%%%%%%%%%%%%%%%%%%%%%%%%%%%
\section{Approximate dynamics: classical CD}
%%%%%%%%%%%%%%%%%%%%%%%%%%%%%%%%%%%%%%%%%%%%%%%%%%%%%%%%%%%%%%%%%%%%%%%
\hspace*{\parindent}
Although the GCD identity (Eq. (\ref{21})) is exact
for the case $\hat{B}\in\{\hat{O}^{i}\}$,
the exact computation of the GCD trajectories $A_{\beta}^{c}(t)$
is in general difficult even in simple one-dimensional systems.  
Therefore, approximations to the GCD are necessary,
and we can define various types of approximations.
Among them, two approximations,
the decoupled centroid approximation 
and the classical approximation,
are especially important for practical applications.
The former was introduced by Jang and Voth \cite{JangVoth2},
and several effective classical molecular dynamics of centroid variables
can be derived via this approximation. 
A detailed description of such approximate dynamics
will be provided elsewhere \cite{Horikoshi}.  
On the other hand, the latter, which is presented in this section,
reduces the exact CD 
to various classical dynamics of the path integral beads.

%%%%%%%%%%%%%%%%%%%%%%%%%%%%%%%%%%%%%%%%%%%%%%%
\subsection{Classical approximation}
%%%%%%%%%%%%%%%%%%%%%%%%%%%%%%%%%%%%%%%%%%%%%%%
\hspace*{\parindent}
The classical approximation consists of two steps.
We first impose an assumption
\begin{eqnarray}
A_{\beta}^{c}(t;\{O_{c}^{(i)}\})\simeq
A_{\beta}^{c}(\{O_{c}^{(i)}(t)\}),
\label{22}
\end{eqnarray}
which means that the time-dependent effective classical operator is assumed 
to be a function of dynamical centroid variables $\{O_{c}^{(i)}(t)\}$. 
Then, we approximate dynamical centroid variables by the corresponding centroids
defined by the dynamical path integral beads,
\begin{eqnarray}
O_{c}^{(i)}(t)\simeq O_{0}^{(i)}(t)=
\lim_{P\to\infty}\frac{1}{P}\sum_{j=1}^{P}O^{(i)}(q_{j}(t),p_{j}(t)),
\label{23}
\end{eqnarray}
where the variables $\{q_{j}(t),p_{j}(t)\}$ 
are assumed to evolve in a classical fashion.
This is the classical centroid dynamics (CCD),
an approximate dynamics of the GCD.
Various types of classical time evolutions of the path integral beads
can be introduced to define various types of CCD.
In the CCD, the centroid correlation functions (Eq. (\ref{20})) are evaluated 
by means of the molecular dynamics techniques;
if the dynamics is ergodic,
ensemble averages can be obtained by taking time averages of trajectories
given by solving a set of classical equations of motion.
In the following subsections, we present several variations of the CCD and 
resulting correlation functions.
\\
\hspace*{\parindent}
Here, we give some comments on symmetries of Kubo-transformed correlation functions.
In the case that $(\hat{A},\hat{B})=(A(\hat{q}),B(\hat{q}))$, 
the Kubo-transformed correlation function has the 
same symmetries as the corresponding correlation functions
defined in classical statistical mechanics \cite{CraigManolopoulos}. 
Therefore, in this case, the centroid correlation function (Eq. (\ref{20}))
might be calculated 
by means of the classical molecular dynamics techniques.
However, in case of $(\hat{A},\hat{B})=(A(\hat{p}),B(\hat{p}))$,
the time-reversal symmetry of the Kubo-transformed correlation function
holds only if $\hat{A}=\hat{B}$ or the potential
$V(\hat{q})$ is parity symmetric. 
Furthermore, in the case that 
$(\hat{A},\hat{B})=(A(\hat{q}),B(\hat{p}))$ or $(A(\hat{p}),B(\hat{q}))$,
the time-reversal symmetry is in general broken.
These facts suggest that in those cases, 
molecular dynamical calculations of 
centroid correlation functions do not work very much.
%%%%%%%%%%%%%%%%%%%%%%%%%%%%%%%%%%%%%%%%%%%%%%%
\subsection{Phase space CCD}
%%%%%%%%%%%%%%%%%%%%%%%%%%%%%%%%%%%%%%%%%%%%%%%
\hspace*{\parindent}
We first consider a classical dynamics where   
dynamical variables $\{q_{j}(t),p_{j}(t)\}$
evolve according to Hamilton's equations of motion with 
the Hamiltonian $H_{P}^{\rm ps}$ (Eq. (\ref{4})),
\begin{eqnarray}
\frac{d}{dt}q_{j}(t)&=&\frac{\partial H_{P}^{\rm ps}}{\partial p_{j}},
\label{24}\\
\frac{d}{dt}p_{j}(t)&=&-\frac{\partial H_{P}^{\rm ps}}{\partial q_{j}}.
\label{25}
\end{eqnarray}
If the system is ergodic, the correlation function can be evaluated as 
\begin{eqnarray}
\langle B^{c}_{\beta}(0)A^{c}_{\beta}(t)\rangle^{\rm CD}_{\beta}
\simeq\lim_{P\to\infty}
\frac{1}{Z_{\beta}}\prod_{j=1}^{P}
\int\!\!\!\!\int\frac{dq_{j}dp_{j}}{2\pi\hbar}~
e^{-\beta_{P}H_{P}^{\rm ps}}
B^{c}_{\beta}(\{O^{(i)}_{0}(0)\})
A^{c}_{\beta}(\{O^{(i)}_{0}(t)\}).
\label{27}
\end{eqnarray}
We refer to this dynamics as the phase space classical centroid dynamics
(PS-CCD) because the classical time evolution is governed by
the Hamiltonian $H_{P}^{\rm ps}$ 
in the phase space path integral representation. 
\par
It should be noted here that the PS-CCD is not
suited for practical calculations because the imaginary term in the
Hamiltonian, $-ip_{j}(q_{j}-q_{j-1})/(\beta_{P}\hbar)$,
makes the dynamics ill-defined.

%%%%%%%%%%%%%%%%%%%%%%%%%%%%%%%%%%%%%%%%%%%%%%%
\subsection{Fictitious momenta and masses: summary of the PIMD}
%%%%%%%%%%%%%%%%%%%%%%%%%%%%%%%%%%%%%%%%%%%%%%%
\hspace*{\parindent}
The PS-CCD has the problem originating in the fact 
that the Hamiltonian $H_{P}^{\rm ps}$ is complex.
On the other hand, in ordinary PIMD simulations, 
we never encounter this kind of difficulty
because we use the configuration space path integral representation of 
the quantum partition function,
\begin{eqnarray}
Z_{\beta}&=&\lim_{P\to\infty}
\prod_{j=1}^{P}
\sqrt{\frac{m}{2\pi\beta_{P}\hbar^{2}}}
\int dq_{j}~
e^{-\beta_{P}V_{P}^{\rm cs}}.
\label{28}
\end{eqnarray}
This expression can be derived from 
the phase space path integral expression (Eq. (\ref{2b}))
by integrating out the momentum variables $\{p_{j}\}$.
The potential $V_{P}^{\rm cs}$ is given by
\begin{eqnarray}
V_{P}^{\rm cs}(\{q_{j}\})=\sum_{j=1}^{P}\left.\left[~
\frac{1}{2}k_{P}(q_{j}-q_{j-1})^{2}+V(q_{j})~
\right]\right|_{q_{0}=q_{P}},
\label{29}
\end{eqnarray}
where $k_{P}=m/(\beta_{P}^{2}\hbar^{2})$.
Inserting $P$ identities $1=\sqrt{\beta_{P}/(2\pi \tilde{m}_{j})}
\int^{\infty}_{-\infty}d\tilde{p}_{j}~e^{-\beta_{P}\tilde{p}_{j}^{2}/(2\tilde{m}_{j})}$
into $Z_{\beta}$ (Eq. (\ref{28})),
we obtain
\begin{eqnarray}
Z_{\beta}
=\lim_{P\to\infty}\prod_{j=1}^{P}\sqrt{\frac{m}{\tilde{m}_{j}}}
\int\!\!\!\!\int\frac{dq_{j}d\tilde{p}_{j}}{2\pi\hbar}~
e^{-\beta_{P}H_{P}^{\rm cs}},
\label{28a}
\end{eqnarray}
where the Hamiltonian is given by
\begin{eqnarray} 
H_{P}^{\rm cs}=\sum_{j=1}^{P}\frac{\tilde{p}_{j}^{2}}{2\tilde{m}_{j}}
+V_{P}^{\rm cs}.
\label{28b}
\end{eqnarray}
New momenta $\{\tilde{p}_{j}\}$ and masses $\{\tilde{m}_{j}\}$ 
are introduced as fictitious momenta and masses, respectively.
This is the partition function 
used in the configuration space PIMD simulations 
\cite{BerneThirumalai,BerneCiccottiCoker}.
\\
\hspace*{\parindent}
In most of PIMD simulations,
we often use a transformation from the bead variables $\{q_{j}\}$ to 
normal modes $\{x_{n}\}$,
\begin{eqnarray}
x_{n}=\sum^{P}_{j=1}U_{nj}q_{j},
\label{29a}
\end{eqnarray}
where $U$ is a unitary matrix.
If we transform the variables $\{q_{j}\}$ in Eq. (\ref{28})
using an orthogonal matrix 
\begin{eqnarray}
U_{nj}=\frac{1}{\sqrt{P}}\left(\cos\frac{2\pi nj}{P}-\sin\frac{2\pi nj}{P}
\right),
\label{29b}
\end{eqnarray}
and insert $P$ identities $1=\sqrt{\beta_{P}/(2\pi \tilde{m}_{n})}
\int^{\infty}_{-\infty}d\tilde{p}_{n}~e^{-\beta_{P}\tilde{p}_{n}^{2}/(2\tilde{m}_{n})}$
into Eq. (\ref{28}),
we obtain another expression of the quantum partition function,
\begin{eqnarray}
Z_{\beta}
=\lim_{P\to\infty}\prod_{n=1}^{P}\sqrt{\frac{m}{\tilde{m}_{n}}}
\int\!\!\!\!\int\frac{dx_{n}d\tilde{p}_{n}}{2\pi\hbar}~
e^{-\beta_{P}H_{P}^{\rm NM}},
\label{29e}
\end{eqnarray}
where the Hamiltonian is given by
\begin{eqnarray}
H_{P}^{\rm NM}(\{x_{n},\tilde{p}_{n}\})=\sum_{n=1}^{P}\left[
\frac{\tilde{p}_{n}^{2}}{2\tilde{m}_{n}}+
\left(2k_{P}\sin^{2} \frac{\pi n}{P}\right) x_{n}^{2}
+\tilde{V}(\{x_{n}\})
\right].
\label{29d}
\end{eqnarray}
This kind of expression of the partition function 
is used in the normal mode PIMD
simulations \cite{BerneThirumalai,BerneCiccottiCoker}.
%%%%%%%%%%%%%%%%%%%%%%%%%%%%%%%%%%%%%%%%%%%%%%%
\subsection{Configuration space CCD}
%%%%%%%%%%%%%%%%%%%%%%%%%%%%%%%%%%%%%%%%%%%%%%%
\hspace*{\parindent}
Here we introduce another type of CCD,
the configuration space classical centroid dynamics (CS-CCD).
There are two expressions of the CS-CCD.
One is a classical dynamics governed by Hamilton's equations of motion
for the bead variables $\{q_{j}(t),\tilde{p}_{j}(t)\}$,
\begin{eqnarray}
\frac{d}{dt}q_{j}(t)&=&\frac{\partial H_{P}^{\rm cs}}{\partial \tilde{p}_{j}},
\label{31}\\
\frac{d}{dt}\tilde{p}_{j}(t)&=&-\frac{\partial H_{P}^{\rm cs}}{\partial q_{j}},
\label{32}
\end{eqnarray}
where $H_{P}^{\rm cs}$ is the Hamiltonian (Eq. (\ref{28b})) used 
in the configuration space PIMD.
If the dynamics is ergodic,
the correlation function is given by
\begin{eqnarray}
\langle B^{c}_{\beta}(0)A^{c}_{\beta}(t)\rangle^{\rm CD}_{\beta}
\simeq\lim_{P\to\infty}
\frac{1}{Z_{\beta}}\prod_{j=1}^{P}\sqrt{\frac{m}{\tilde{m}_{j}}}
\int\!\!\!\!\int\frac{dq_{j}d\tilde{p}_{j}}{2\pi\hbar}~
e^{-\beta_{P}H_{P}^{\rm cs}}
B^{c}_{\beta}(\{O^{(i)}_{0}(0)\})A^{c}_{\beta}(\{O^{(i)}_{0}(t)\}).
\label{30}
\end{eqnarray}
We refer to this dynamics as 
the ring polymer CS-CCD 
because the variables $\{q_{j}(t)\}$ form a ring polymer
where the nearest neighbors are connected with the spring constant $k_{P}$.
\\
\hspace*{\parindent}
The other expression of the CS-CCD is a classical dynamics of 
the normal mode variables $\{x_{n}(t),\tilde{p}_{n}(t)\}$. 
This is the normal mode CS-CCD,
which is governed by Hamilton's equations of motion
with the Hamiltonian $H_{P}^{\rm NM}$ (Eq. (\ref{29d})),
\begin{eqnarray}
\frac{d}{dt}x_{n}(t)&=&\frac{\partial H_{P}^{\rm NM}}{\partial \tilde{p}_{n}},
\label{31b}\\
\frac{d}{dt}\tilde{p}_{n}(t)&=&-\frac{\partial H_{P}^{\rm NM}}{\partial x_{n}}.
\label{32b}
\end{eqnarray}
If the ergodicity of the dynamics is satisfied,
the correlation function is given by 
\begin{eqnarray}
\langle B^{c}_{\beta}(0)A^{c}_{\beta}(t)\rangle^{\rm CD}_{\beta}
\simeq\lim_{P\to\infty}
\frac{1}{Z_{\beta}}\prod_{n=1}^{P}
\sqrt{\frac{m}{\tilde{m}_{n}}}
\int\!\!\!\!\int\frac{dx_{n}d\tilde{p}_{n}}{2\pi\hbar}~
e^{-\beta_{P}H_{P}^{\rm NM}}
B^{c}_{\beta}(\{O^{(i)}_{0}(0)\})A^{c}_{\beta}(\{O^{(i)}_{0}(t)\}).
\label{32a}
\end{eqnarray}
\hspace*{\parindent}
Here, we give two comments on the CS-CCD:
(1) In general systems, 
the CS-CCD is exact only for the calculation of the static correlation functions
$\langle B^{c}_{\beta}(0)A^{c}_{\beta}(0)\rangle$ for position-dependent operators
$(A(\hat{q}),B(\hat{q}))$.
In the case of $t\ne 0$ or momentum-dependent operators,
the fictitious momenta $\{\tilde{p}_{j}\}$ (or $\{\tilde{p}_{n}\}$)
enlarge the gap between the CS-CCD correlation function and 
the Kubo-transformed correlation function.   
Therefore, the validity of 
the CS-CCD expressions (Eqs. (\ref{30}) or (\ref{32a}))
should be checked depending on the situation.
(2) The choice of the fictitious masses is crucial.
In the normal mode CS-CCD, if the fictitious mass of $P$-th normal mode 
is chosen as $\tilde{m}_{P}=m$,
then the CS-CCD gives the exact correlation function
$\langle B^{c}_{\beta}(0)A^{c}_{\beta}(0)\rangle$ for 
$\hat{A}=\hat{B}=\hat{p}\in\{\hat{O}^{(i)}\}$.
This is because the choice $\tilde{m}_{P}=m$
connects the fictitious momentum $\tilde{p}_{P}$ with
the physical momentum centroid $p_{0}$ via the centroid expression of
the density (Eqs. (\ref{10b}) and (\ref{11})).
In the ring polymer CS-CCD, this condition can be satisfied
by setting $\{\tilde{m}_{j}=m\}$, i.e. setting 
all fictitious masses equal to the physical mass.
%%%%%%%%%%%%%%%%%%%%%%%%%%%%%%%%%%%%%%%%%%%%%%%
\subsection{Ring polymer molecular dynamics}
%%%%%%%%%%%%%%%%%%%%%%%%%%%%%%%%%%%%%%%%%%%%%%%
\hspace*{\parindent}
Finally, we present a special case of the ring polymer CS-CCD.
If the operators $(\hat{A},\hat{B})$ belong to the operator set $\{\hat{O}^{(i)}\}$,
and all fictitious masses are chosen to be the physical mass,
$\{\tilde{m}_{j}=m\}$, 
the ring polymer CS-CCD correlation function (Eq. (\ref{30})) becomes
\begin{eqnarray}
\langle B^{c}_{\beta}(0)A^{c}_{\beta}(t)\rangle^{\rm CD}_{\beta}
\simeq\lim_{P\to\infty}
\frac{1}{Z_{\beta}}\prod_{j=1}^{P}
\int\!\!\!\!\int\frac{dq_{j}d\tilde{p}_{j}}{2\pi\hbar}~
e^{-\beta_{P}H_{P}^{\rm cs}}B_{0}(0)A_{0}(t).
\label{33}
\end{eqnarray} 
This dynamics is equivalent to the RPMD
proposed by Craig and Manolopoulos \cite{CraigManolopoulos}.
As we mentioned in the preceding subsection,  
the RPMD reproduces the exact correlation functions at $t=0$
for $(\hat{A},\hat{B})=(A(\hat{q}),B(\hat{q}))$ or $(\hat{p},\hat{p})$.
Recently, it has been shown that 
for the calculations of 
$\langle \hat{q}(0)\hat{q}(t)\rangle^{\rm K}_{\beta}$
and $\langle \hat{p}(0)\hat{p}(t)\rangle^{\rm K}_{\beta}$,
the RPMD is correct up to 
$O(t^{6})$ and $O(t^{4})$,
respectively \cite{BraamsManolopoulos}.
%%%%%%%%%%%%%%%%%%%%%%%%%%%%%%%%%%%%%%%%%%%%%%%
\subsection{Simple example: a harmonic system}
%%%%%%%%%%%%%%%%%%%%%%%%%%%%%%%%%%%%%%%%%%%%%%%
\hspace*{\parindent}
Here, we consider the case $\hat{A}=\hat{B}=\hat{q}^{2}\in \{\hat{O}^{(i)}\}$,
and show the results of two types of approximate CD:
the RPMD and the normal mode CS-CCD.
We consider a simple system with a harmonic potential
$V(\hat{q})=(m\omega^{2}/2)\hat{q}^{2}$.
In this system, the exact Kubo-transformed 
``position squared'' autocorrelation function is given by
\begin{eqnarray}
\langle \hat{q}^{2}(0)\hat{q}^{2}(t)\rangle_{\beta}^{\rm K}
&=&\frac{\hbar^2}{4m^{2}\omega^{2}}\left[
\frac{2}{\beta\hbar\omega}
\coth\frac{\beta\hbar\omega}{2}
\cos 2\omega t
+2\coth^2\frac{\beta\hbar\omega}{2}-1\right]
.\label{40}
\end{eqnarray}
The corresponding RPMD correlation function is calculated as 
\begin{eqnarray}
\langle \hat{q}^{2}(0)\hat{q}^{2}(t)\rangle_{\beta}^{\rm RPMD}
&=&\lim_{P\to\infty}
\frac{1}{\beta^{2}m^{2}}\left[
\sum_{n=1}^{P}\frac{1}{\omega_{n}^{4}}(\cos 2\omega_{n} t +1)+
\sum_{n=1}^{P}\sum_{l=1}^{P}\frac{1}{\omega_{n}^{2}\omega_{l}^{2}}
\right]
,\label{41}
\end{eqnarray}
where $\omega_{n}=\sqrt{\omega^{2}+(4k_{P}/m)\sin^{2}(\pi n/P)}$.
The corresponding normal mode CS-CCD correlation function is also
obtained as 
\begin{eqnarray}
\langle \hat{q}^{2}(0)\hat{q}^{2}(t)\rangle_{\beta}^{\rm NM}
&=&\lim_{P\to\infty}
\frac{1}{\beta^{2}m^{2}}\left[
\sum_{n=1}^{P}\frac{1}{\omega_{n}^{4}}(\cos 2\Omega_{n} t +1)+
\sum_{n=1}^{P}\sum_{l=1}^{P}\frac{1}{\omega_{n}^{2}\omega_{l}^{2}}
\right]
,\label{42}
\end{eqnarray}
where $\Omega_{n}=\omega_{n}\sqrt{m/\tilde{m}_{n}}$.
\\
\hspace*{\parindent}
Figure \ref{fig:Fig1} shows the plot of the three 
correlation functions, Eqs. (\ref{40})--(\ref{42}),
at two different temperatures
with the parameters $\hbar=k_{B}=m=\omega=1$.
We set the number of beads $P$ as $1000$, 
which is so large as to make the results converged sufficiently.  
The fictitious masses of the normal modes are chosen as
$\{\tilde{m}_{n}=m+(4m/\beta_{P}^{2}\hbar^{2}\omega^{2})\sin^{2}(\pi n/P)\}$.
This choice makes each frequency $2 \Omega_{n}$ equal to $2\omega$.
At the higher temperature $\beta=1$ (Fig. \ref{fig:Fig1} (a)),  
the RPMD correlation function and the normal mode CS-CCD correlation function 
are almost identical, and they coincide with 
the exact Kubo-transformed correlation function at $t=0$.
However, as the time $t$ increases, they slightly deviate from the exact
correlation function.
These deviations become more significant 
at the lower temperature $\beta=10$ (Fig. \ref{fig:Fig1} (b)).
As is observed clearly in Fig. \ref{fig:Fig1} (b),
the RPMD correlation function damps with time.      
This is because the mode summation $\sum_{n=1}^{P}$
causes a dephasing effect.
The lower the temperature falls,
the more modes with different frequencies 
are relevant in the summation $\sum_{n=1}^{P}$.
Therefore, the dephasing effect becomes stronger at lower temperature.
On the other hand,
the normal mode CS-CCD correlation function is free from 
such a dephasing effect,
thanks to the special choice of fictitious masses $\{\tilde{m}_{n}\}$.
However, as one can also see in Fig. \ref{fig:Fig1} (b),
the short time behavior of 
$\langle \hat{q}^{2}(0)\hat{q}^{2}(t)\rangle_{\beta}^{\rm NM}$
is worse than
$\langle \hat{q}^{2}(0)\hat{q}^{2}(t)\rangle_{\beta}^{\rm RPMD}$.
This is because the mass choice adopted in the RPMD, $\{\tilde{m}_{j}=m\}$,
is the optimized one 
to reproduce the short time behaviors
of exact Kubo-transformed correlation functions \cite{BraamsManolopoulos}.
These observations in the simple harmonic system show 
the importance of the mass choice.
If you focus on the short time behaviors of correlation functions, 
you should choose $\{\tilde{m}_{j}=m\}$.
On the other hand, if you respect the dynamical quantum effects 
such as coherent oscillations of correlation functions, 
another choice might be better.
%%%%%%%%%%%%%%%%%%%%%%%%%%%%%%%%%%%%%%%%%%%%%%%%%%%%%%%%%%%%%%%%%%%%%%%
\section{Concluding remarks}
\begin{figure}
\begin{tabular}{c}
\includegraphics[width=130mm]{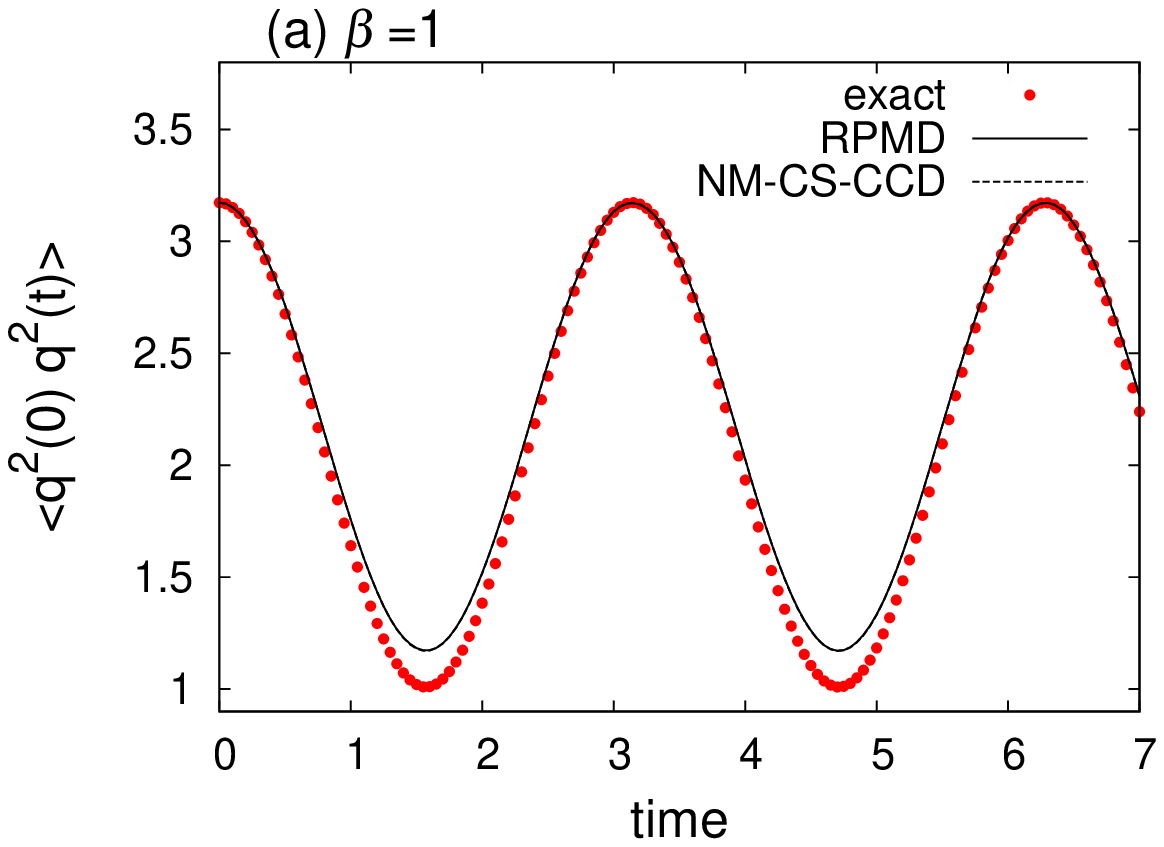}\\
\includegraphics[width=130mm]{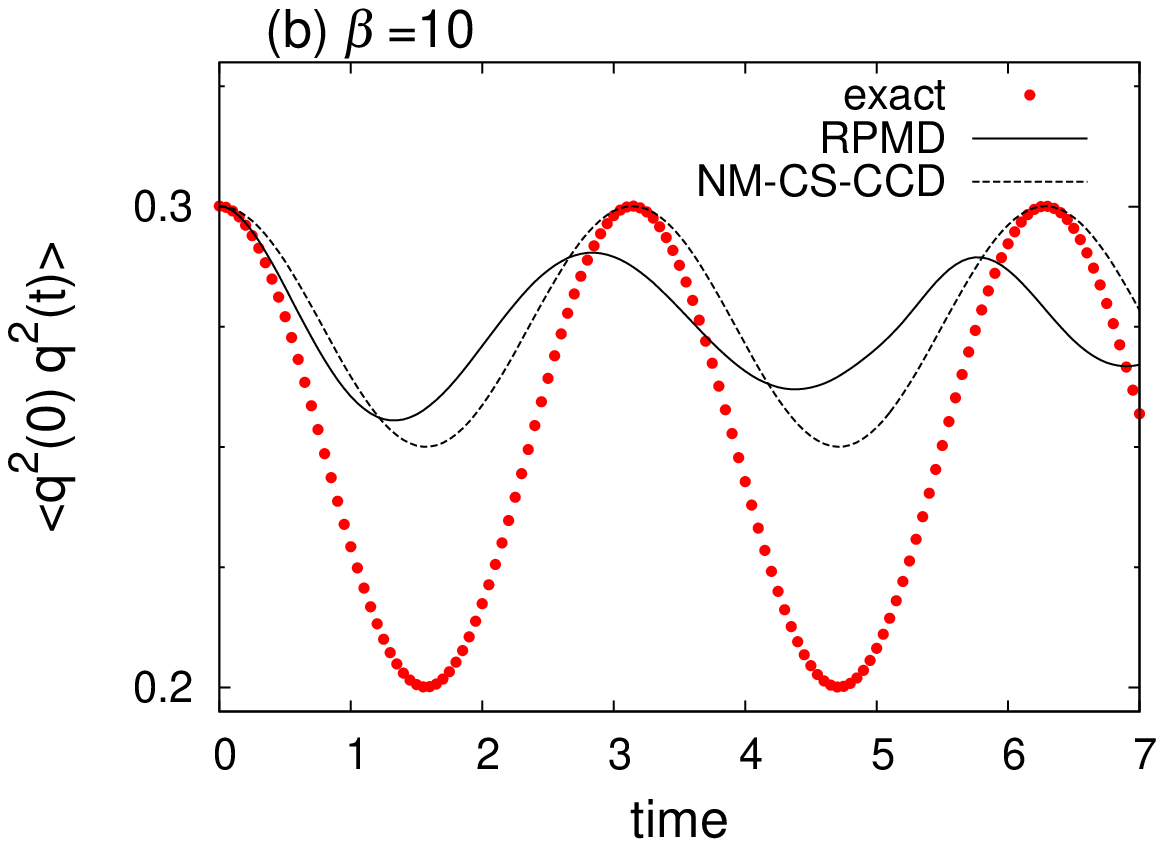}\\
\end{tabular}
\vspace{0mm}
\caption{
The plot of the exact Kubo-transformed correlation function (Eq. (\ref{40})),
the RPMD correlation function (Eq. (\ref{41})),
and the normal mode CS-CCD correlation function (Eq. (\ref{42}))
for the quantum harmonic oscillator:
(a) in the higher temperature regime ($\beta=1$) and 
(b) in the lower temperature regime ($\beta=10$). 
}
\label{fig:Fig1}
\end{figure}

\begin{figure}[htb]
{\Large
\begin{center}
\setlength{\unitlength}{1mm}
\begin{picture}(160,180)
\put(25,157){\framebox(110,23){%
 \shortstack{Kubo-transformed correlation functions\\ \\
$\langle \hat{B}(0)\hat{A}(t)\rangle^{\rm K}_{\beta}$
}}}
\put(25,110){\framebox(110,27){%
 \shortstack{centroid correlation functions\\ \\
$\langle B^{c}_{\beta}(0) A^{c}_{\beta}(t)\rangle^{\rm CD}_{\beta}$\\ \\
by the generalized centroid dynamics
}}}
\put(5,64){\framebox(74,26){%
 \shortstack{phase space \\ \\
classical centroid dynamics
}}}
\put(98,52){\framebox(60,25){%
 \shortstack{generalized centroid\\ \\
             molecular dynamics
}}}
\put(5,38){\framebox(74,26){%
 \shortstack{configuration space \\ \\
classical centroid dynamics             
}}}
\put(12,-7){\framebox(60,25){%
 \shortstack{ring polymer \\ \\
             molecular dynamics \\
{\normalsize Craig and Manolopoulos (2004)}
}}}
\put(98,-7){\framebox(60,25){%
\shortstack{centroid  \\ \\
             molecular dynamics\\
{\normalsize Cao and Voth (1994)}
}}}
\put(80,147){\thicklines\vector(0,-1){10}}
\put(80,147){\thicklines\vector(0,1){10}}
\put(42,110){\thicklines\vector(0,-1){20}}
\put(118,110){\thicklines\vector(0,-1){33}}
\put(42,38){\thicklines\vector(0,-1){20}}
\put(118,52){\thicklines\vector(0,-1){34}}
\put(85,150){\shortstack{equivalent}}
\put(85,141){\shortstack{if $\hat{B}\in\{\hat{O}^{(i)}\}$}}
\put(47,102){\shortstack{classical}}
\put(47,96){\shortstack{approximation}}
\put(123,98){\shortstack{decoupled}}
\put(123,92){\shortstack{centroid}}
\put(123,86){\shortstack{approximation}}
\put(47,30){\shortstack{$\hat{A},\hat{B}\in\{\hat{O}^{(i)}\}$}}
\put(47,22){\shortstack{$\{\tilde{m}_{j}=m\}$}}
\put(121,34){\shortstack{$\{\hat{O}^{(i)}\}=\{\hat{q},\hat{p}\}$}}
\end{picture}
\end{center}
}
\vspace{30mm}
\caption{
Schematic diagram for the GCD and its approximate dynamics. 
The first up-down arrow represents the equivalence between 
Kubo-transformed correlation functions and 
centroid correlation functions given by the GCD (Eq. (\ref{21})). 
The left-hand route to the RPMD is 
obtained via the classical approximation.
Detailed descriptions on the right-hand route to the CMD 
have not been presented in this article \cite{Horikoshi}. 
}
\label{fig:Fig2}
\end{figure}
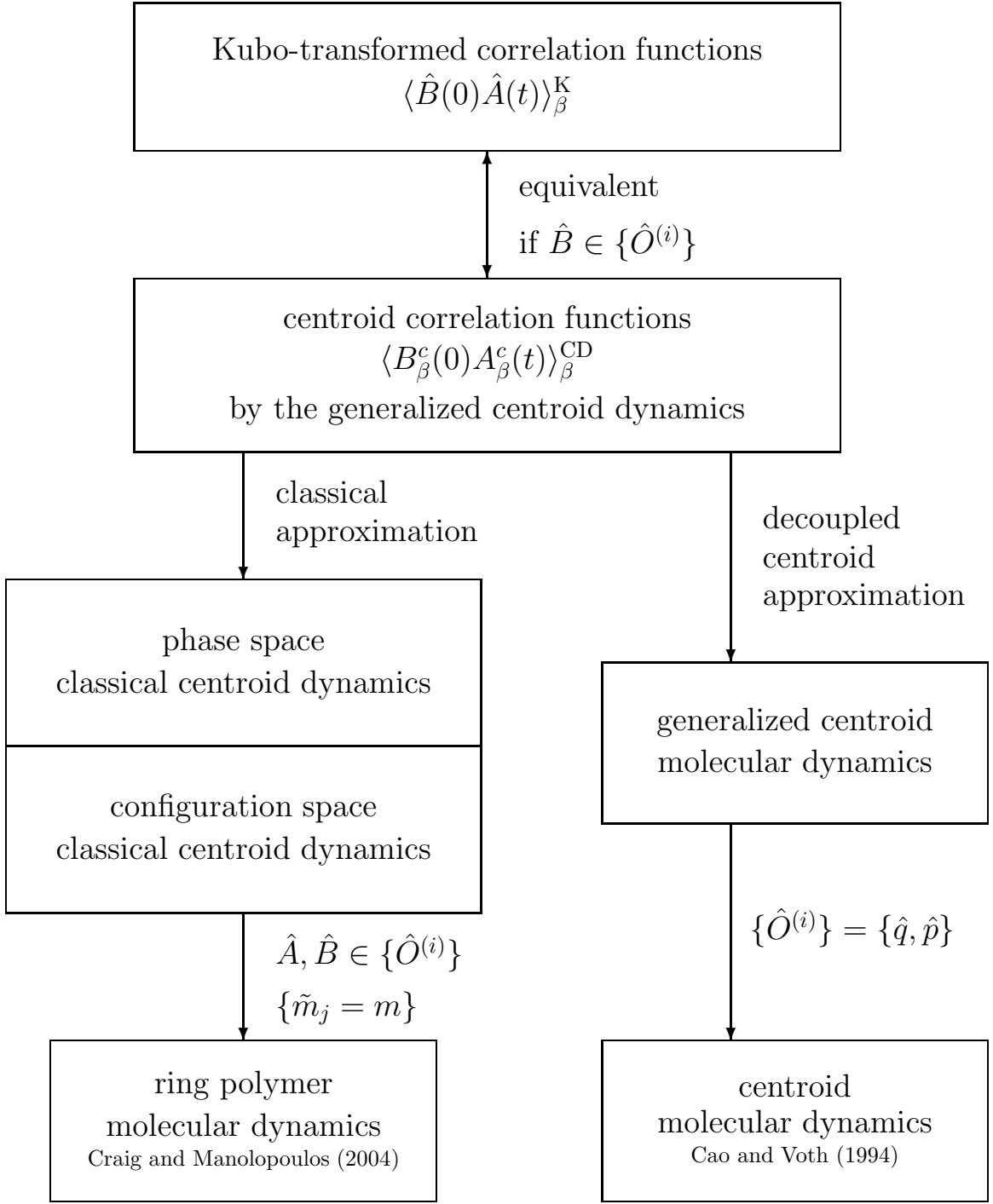

\hspace*{\parindent}
In this work, we have developed
an exact CD for composite operators of $\hat{q}$ and $\hat{p}$,
the GCD, which gives 
an exact identity between a centroid correlation function and
a Kubo-transformed correlation function (Eq. (\ref{21})).
We have then proposed the classical approximation
and the corresponding approximate GCD, the CCD (Eqs. (\ref{22}) and (\ref{23})).
Introducing several types of classical time evolutions,
we have defined several types of CCD,
the PS-CCD (Eqs. (\ref{24}) and (\ref{25})),
the ring polymer CS-CCD (Eqs. (\ref{31}) and (\ref{32})),
and  the normal mode CS-CCD (Eqs. (\ref{31b}) and (\ref{32b})).
If we consider operators $\hat{A}$ and $\hat{B}$
which belong to the operator set $\{\hat{O}^{(i)}\}$, 
and set the fictitious masses equal to the physical mass, 
$\{\tilde{m}_{j}=m\}$,
the ring polymer CS-CCD becomes equivalent to the RPMD
proposed by Craig and Manolopoulos (Eq. (\ref{33})) \cite{CraigManolopoulos}. 
A schematic diagram of various approximate dynamics is given in Fig. \ref{fig:Fig2}.
The results of simple calculations in a harmonic system
have shown that the choice of fictitious masses is 
crucial in CS-CCD calculations.
The RPMD might be the best CS-CCD method to calculate 
correlation functions in the short time region.
\\
\hspace*{\parindent}
We have shown that the RPMD can be formulated as an approximate GCD.
However, the classical approximation employed there is rather crude.
The physical meaning of the approximation should be clarified,
and systematic schemes to improve the approximation
should be developed.
Recently, a relationship between the RPMD and the semiclassical instanton theory
has been discussed in the deep tunneling regime,
and it has been shown that the RPMD can be systematically improved 
in that regime \cite{RichardsonAlthorpe}. 
\\
\hspace*{\parindent}
Finally, we briefly mention another approximation scheme. 
Applying the decoupled centroid approximation to the GCD,
we obtain the generalized centroid molecular dynamics (GCMD) \cite{Horikoshi}.
In the case of $N=2$ and 
$(\hat{O}^{(1)},\hat{O}^{(2)})=(\hat{q},\hat{p})$,
the GCMD is reduced to the CMD proposed by Cao and Voth \cite{CaoVoth}.
The relations between these methods are summarized in Fig. \ref{fig:Fig2}.
The GCMD is a dynamics of centroid variables $\{O_{c}^{(i)}(t)\}$
rather than centroids $\{O_{0}^{(i)}(t)\}$ 
defined by the summation of the path integral beads.
Therefore, the GCMD correlation functions 
are expected to be free from the summation-induced dephasing effect 
seen in the RPMD correlation functions (Fig. \ref{fig:Fig1}).
However, in general, it is not so easy to formulate the GCMD and to implement it 
for practical calculations.
This is because the GCMD is a non-Hamiltonian dynamics which 
should be formulated in an extended phase space spanned 
by canonical multiplets \cite{Horikoshi}. 
A proper formulation of the GCMD might be given by
a generalized Hamiltonian dynamics proposed by Nambu \cite{Nambu}.

%%%%%%%%%%%%%%%%%%%%% Acknowledgments  %%%%%%%%%%%%%%%%%%%%%%%%%%%%%%%%

%%%%%%%%%%%%%%%%%%%%% References  %%%%%%%%%%%%%%%%%%%%%%%%%%%%%%%%%%%%%   

\newpage

\newpage

\end{document}